\documentclass[%
 aip,
 amsmath,amssymb,
 reprint,%
]{revtex4-1}

\usepackage{graphicx}% Include figure files
\usepackage{dcolumn}% Align table columns on decimal point
\usepackage{bm}% bold math
\usepackage[utf8]{inputenc}
\usepackage[T1]{fontenc}
\usepackage{mathptmx}
\usepackage{etoolbox}
\usepackage{color}
\usepackage{txfonts}

\newcommand{\mk}[1]{{\textcolor{black}{#1}}}
\newcommand{\cyan}[1]{\textcolor{black}{#1}}

\makeatletter
\def\@email#1#2{%
 \endgroup
 \patchcmd{\titleblock@produce}
  {\frontmatter@RRAPformat}
  {\frontmatter@RRAPformat{\produce@RRAP{*#1\href{mailto:#2}{#2}}}\frontmatter@RRAPformat}
  {}{}
}%
\makeatother
\begin{document}

\preprint{AIP/123-QED}

\title[Materials design of nickelate superconductors]{{\it Ab initio} Materials Design of Superconductivity in $d^9$ Nickelates}

\author{Motoharu Kitatani}
\affiliation{ 
RIKEN Center for Emergent Matter Science, 2-1 Hirosawa, Wako, Saitama 351-0198, Japan 
}
\affiliation{Department of Material Science, University of Hyogo, Ako, Hyogo 678-1297, Japan}
\email{kitatanimo@gmail.com}
\author{Yusuke Nomura}
\affiliation{ 
Department of Applied Physics and Physico-Informatics, Keio University, 3-14-1 Hiyoshi, Kohoku-ku, Yokohama, 223-8522, Japan
}

\author{Motoaki Hirayama}
\affiliation{ 
RIKEN Center for Emergent Matter Science, 2-1 Hirosawa, Wako, Saitama 351-0198, Japan 
}
\affiliation{
Quantum-Phase Electronics Center, The University of Tokyo,7-3-1 Hongo, Bunkyo-ku, Tokyo 113-8656, Japan
}

\author{Ryotaro Arita}
\affiliation{ 
RIKEN Center for Emergent Matter Science, 2-1 Hirosawa, Wako, Saitama 351-0198, Japan 
}
\affiliation{
Research Center for Advanced Science and Technology, University of Tokyo
4-6-1, Komaba, Meguro-ku, Tokyo 153-8904, Japan
}

\date{\today}% It is always \today, today,

\begin{abstract}
Motivated by the recent theoretical materials design of superconducting $d^9$ nickelates for which the charge transfer from the NiO$_2$ to the block layer is completely suppressed [M. Hirayama {\it et al.}, Phys. Rev. B {\bf 101}, 075107 (2020)], we perform a calculation based on the dynamical vertex approximation and obtain the phase diagram of RbCa$_2$NiO$_3$ and $A_2$NiO$_2$Br$_2$ where $A$ is a cation with a valence of 2.5+. 
%Since the onsite Hubbard $U$ of these nickelates is less screened and significantly large, the superconducting transition temperature is lower than those of the infinite-layer nickelates such as NdNiO$_2$ or the quintuple-layer nickelate Nd$_6$Ni$_5$O$_{12}$. 
We show that the phase diagram of these nickelates exhibits the same essential features as those found in cuprates. 
Namely, superconductivity appears upon hole-doping into an antiferromagnetic Mott insulator, and the superconducting transition temperature shows a dome-like shape.  
This demonstrates that the electron correlations play an essential role in nickelate superconductors 
and we can control them by changing block layers.
%However, these nickelates may provide us with a unique opportunity to study the electron-hole asymmetry of the Mott insulating states in the nickelate superconductors.
\end{abstract}

\maketitle

\section{Introduction}
Unconventional superconductivity in strongly correlated electron systems such as the cuprates~\cite{Bednorz_1986}, iron-based superconductors~\cite{Kamihara_2008},
Sr$_2$RuO$_4$~\cite{Maeno_1994}, and Na$_x$CoO$_2\cdot y$H$_2$O~\cite{Takada_2003} has been one of the central issues in condensed matter physics. While we still lack a generally accepted explanation for their pairing mechanism, there is one characteristic common feature in these compounds: They have a crystal structure with stacked two-dimensional layers. For such layered materials, various families have been synthesized by inserting different layers or changing the stacking patterns. 

It is of great interest to note that the recently discovered $d^9$ nickelate superconductors~\cite{Li_2019,D_Li_2020,Osada_2020,Osada_2020_2,Osada_2021,SW_Zeng_2020,SW_Zeng_2022} $R$NiO$_2$ ($R$=La, Pr, Nd) also share this common feature (for recent reviews, see, e.g.,  Refs.~\onlinecite{Norman_2020,Pickett_2021,J_Zhang_2021,Botana_2021,Y_Ji_2021,Nomura_2022,Q_Gu_2022,Chow_2022,X_Zhou_2022}). Namely, they consist of the NiO$_2$ layer and block layer, forming the infinite-layer structure.  
The maximum transition temperature $T_{\rm c}$ is about 15 K~\cite{Li_2019,D_Li_2020,Osada_2020,Osada_2020_2,Osada_2021,SW_Zeng_2020,SW_Zeng_2022}. 
\mk{It is further enhanced to $20\sim30$K by changing the substrate or applying pressure~\cite{Ren_2021,NN_Wang_2021,Lee_2022}.}
A theoretical phonon calculation has shown that the electron-phonon coupling is too weak to explain the experimental $T_{\rm c}$~\cite{Nomura_2019}.
Thus, the conventional mechanism is unlikely, and various unconventional mechanisms have been discussed theoretically~\cite{Sakakibara_2020,Hirsch_2019,Wu_2020,Werner_2020,YH_Zhang_2020,J_Chang_2020,Kitatani_2020,Adhikary_2020,Z_Wang_2020,C_Lu_2022,TY_Xie_arXiv,Karp_arXiv,M_Jiang_arXiv,Kreisel_arXiv}. 
On the experimental side, all the reports so far are consistent in that the pairing symmetry is not a simple $s$-wave, although there exists a discrepancy in the proposed symmetry~\cite{Q_Gu_2020,Chow_arXiv_2,Harvey_arXiv}.

More recently, a quintuple-layer nickel superconductor Nd$_6$Ni$_5$O$_{12}$ has been synthesized~\cite{Pan_2021}.
%and is attracting much attention. 
This has proven that superconductivity can be realized in nickelates other than infinite-layer compounds. 
Therefore, it is interesting to consider different layered structures and think of manipulating the electronic structure of nickelate superconductors.
%Exploiting this feature, one can think of manipulating the electronic structure of the nickelate superconductors. 
Indeed, a variety of theoretical materials design has been performed~\cite{Hirayama_2020,Kitamine_2020,Pardo_arXiv}. In particular, it has been shown that we can control the charge transfer from the NiO$_2$ layer to the block layer~\cite{Hirayama_2020}.

\begin{figure}
\centering
\includegraphics[width=8cm,bb=0 0 324 395]{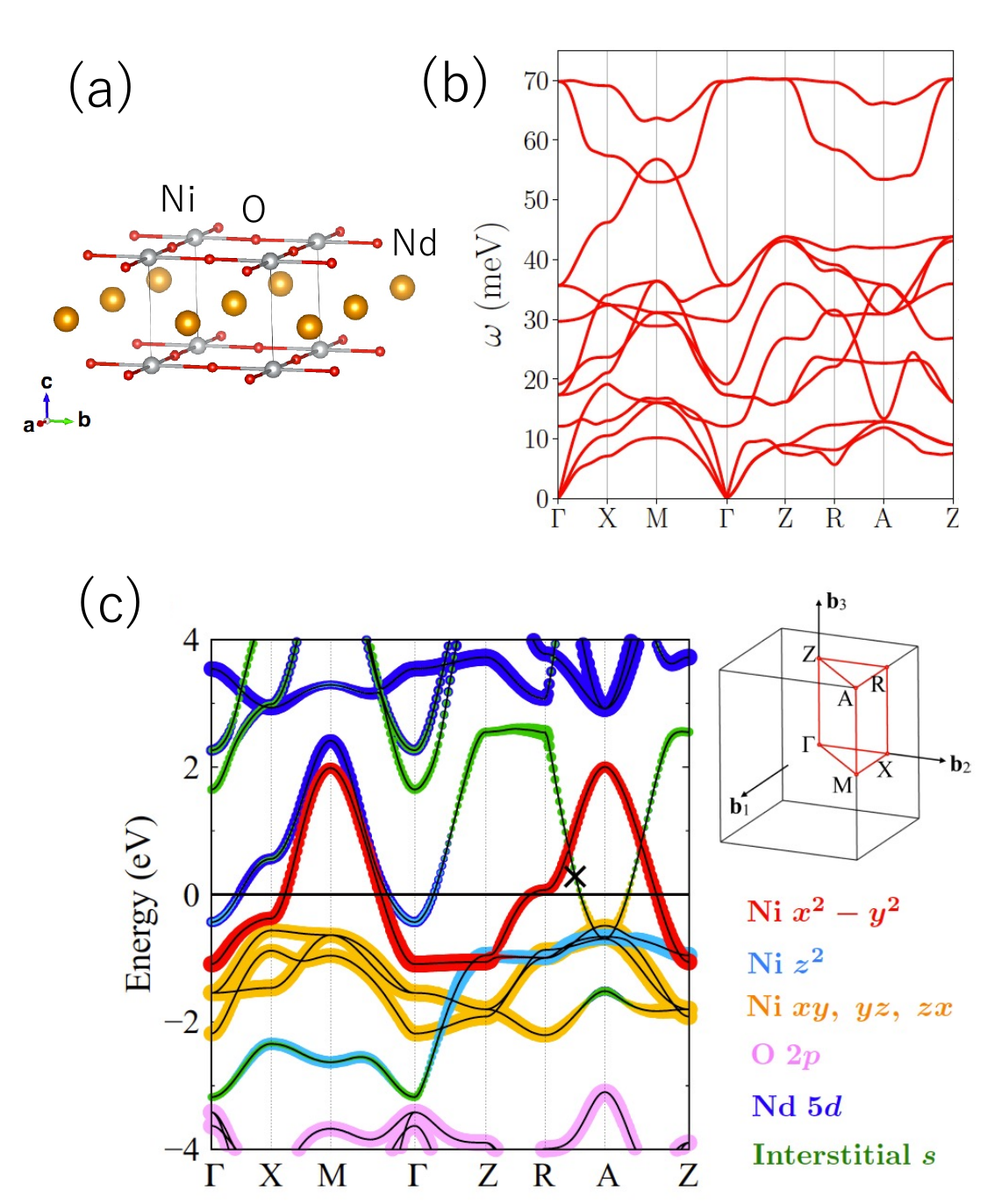}
\caption{(a) Crystal structure, (b) phonon dispersion, and (c) electronic structure of NdNiO$_2$. Taken from Ref.~\onlinecite{Nomura_2019} and Ref.~\onlinecite{Nomura_2022}.}
\label{Fig1}
\end{figure}

In Fig.~\ref{Fig1}, we show the crystal structure, phonon dispersion, and electronic band dispersion of NdNiO$_2$~\cite{Nomura_2019,Nomura_2022}. We see that the electronic structure of NdNiO$_2$ is similar to that of the cuprates in that only the 3$d_{x^2-y^2}$ orbital among the five 3$d$ orbitals contribute to the formation of the Fermi surface. The dispersion can be represented by a simple tight-binding model on the two-dimensional square lattice. However, there are also several distinct differences between NdNiO$_2$ and the cuprates: In NdNiO$_2$, the oxygen 2$p$ level is far below the Fermi level, and the system belongs to the Mott-Hubbard regime~\cite{Hepting_2020,Fu_arXiv,Goodge_2021} in the Zaanen-Sawatzky-Allen phase diagram~\cite{Zaanen_1985}. Thus the hybridization between the oxygen 2$p$ and Ni 3$d$ orbitals is weaker than that in the cuprates which reside in the charge-transfer regime. In addition, there is substantial charge transfer from the NiO$_2$ layer to the block layer~\cite{Lee_2004,Botana_2020}, and the Nd 5$d$ states and interstitial $s$ state form Fermi pockets around the $\Gamma$ and $A$ points.

In Ref.~\onlinecite{Hirayama_2020}, three of the present authors have shown that there are dynamically stable layered $d^9$ nickelates for which the charge transfer from the NiO$_2$ layer to the block layer is completely suppressed. There, only the Ni 3$d_{x^2-y^2}$ orbital forms a two-dimensional Fermi surface, whose shape is very similar to that of the cuprates. For these nickelates, effective low-energy models in terms of the Wannier functions~\cite{Pizzi_2020} for the Ni 3$d_{x^2-y^2}$ orbital and O 2$p$ orbitals have been derived using the constrained random phase approximation (cRPA)~\cite{Aryasetiawan_2004,Nakamura_2021}. In Ref.~\onlinecite{Nomura_2020}, the magnetic exchange coupling $J$ in the NiO$_2$ plane has also been estimated. While it has been known that the cuprate superconductors have a large exchange coupling of $\sim$140 meV~\cite{Lee_2006}, these nickelates also have a significant exchange coupling as large as 80-100 meV\footnote{
We note that the strength of the magnetic exchange coupling $J$ in the infinite-layer nickelates is controversial. 
Experimentally, Ref.~\onlinecite{Fu_arXiv} (Raman experiment using NdNiO$_2$ bulk samples) and Ref.~\onlinecite{Lu_2021} (resonant inelastic X-ray scattering experiment for NdNiO$_2$ thin film samples) gave $J=25$ meV and $J=64(3)$ meV, respectively. 
There is no agreement in theoretical estimates, either~\cite{Jiang_2020,Ryee_2020,HuZhang_2020,GM_Zhang_2020,Z_Liu_2020,Been_2021,Leonov_2020,Leonov_2021,X_Wan_2021,ZJ_Lang_2021,Katukuri_2020,R_Zhang_2021}.
Because the infinite-layer nickelates are not a Mott insulator, an ambiguity exists in the mapping to spin models, which is one of the reasons for the discrepancy in the theoretical estimates~\cite{Nomura_2020}}.

In this paper, we extend these studies and calculate the phase diagram of RbCa$_2$NiO$_3$ and $A_2$NiO$_2$Br$_2$ ($A$ is a cation with a valence of 2.5+) 
by means of the dynamical vertex approximation (D$\Gamma$A)~\cite{Toschi_2007,Katanin_2009,Rohringer_2018}. 
The D$\Gamma$A was also applied to study the phase diagram of the infinite-layer nickelate NdNiO$_2$~\cite{Kitatani_2020} and quintuple-layer nickelate Nd$_6$Ni$_5$O$_{12}$~\cite{Worm_2021}, which allows us to compare the phase diagram among nickelates. 
%For the infinite-layer nickelate NdNiO$_2$~\cite{Kitatani_2020} and quintuple-layer nickelate Nd$_6$Ni$_5$O$_{12}$~\cite{Worm_2021}, the superconducting transition temperature ($T_{\rm c}$) has been calculated using the dynamical vertex approximation (D$\Gamma$A)~\cite{Toschi_2007,Katanin_2009,Rohringer_2018}. 
In RbCa$_2$NiO$_3$ and $A_2$NiO$_2$Br$_2$, differently from the infinite-layer nickelate NdNiO$_2$, superconductivity emerges by a carrier doping into an antiferromagnetic insulator. 
The superconducting transition temperature ($T_{\rm c}$) shows a dome-like shape as a function of doping. 
%We find that the maximum $T_{\rm c}$ of RbCa$_2$NiO$_3$ and $A_2$NiO$_2$Br$_2$ will be lower than those of $R$NiO$_2$ or Nd$_6$Ni$_5$O$_{12}$ since the Hubbard interaction $U$ in the effective model is too large. 
%However, 
Therefore, RbCa$_2$NiO$_3$ and $A_2$NiO$_2$Br$_2$ may provide us with a unique opportunity to study 
the electron-hole asymmetry of the Mott insulating states 
in the nickelate superconductors.    

The structure of the paper is as follows. In Section~\ref{sec:2}, we discuss the phonon dispersion and electronic band dispersion of RbCa$_2$NiO$_3$ and $A_2$NiO$_2$Br$_2$. We show that these nickelates do not have imaginary phonon modes, indicating that they are dynamically stable. Thus although the crystal structures of these nickelates may not be the most stable structure, they are one of the meta-stable structures. Regarding the electronic structure, in contrast with the case of $R$NiO$_2$, the charge transfer from the NiO$_2$ layer to the block layer in RbCa$_2$NiO$_3$ and $A_2$NiO$_2$Br$_2$ is absent. In Section~\ref{sec:3}, we show the results of {\it ab initio} derivation of the effective low-energy model (i.e., the single-orbital Hubbard model) for these nickelates. We see that RbCa$_2$NiO$_3$ and $A_2$NiO$_2$Br$_2$ are more strongly correlated than NdNiO$_2$. In Section~\ref{sec:4}, we solve the single-orbital Hubbard model derived in Section~\ref{sec:3} using the D$\Gamma$A. We obtain the phase diagram for RbCa$_2$NiO$_3$ and $A_2$NiO$_2$Br$_2$ and compare \mk{it} with that of NdNiO$_2$ and Nd$_6$Ni$_5$O$_{12}$. In Section~\ref{sec:5}, we summarize the results obtained in the present study.

\section{Dynamical stability and electronic structure\label{sec:2}}
Let us move on to the materials design of new nickelate superconductors~\cite{Hirayama_2020}. Following the idea for the cuprate superconductors~\cite{Tokura_1990}, four types of block layers and six types of crystal structures for the nickelate superconductors have been proposed. The strategy to suppress the charge transfer from the NiO$_2$ layer to the block layer is the following: If we make the energy level of the block layer sufficiently higher than that of the Ni 3$d_{x^2-y^2}$ orbital, the charge transfer will not occur. Here, it should be noted that the level of Ni 3$d_{x^2-y^2}$ orbital is considerably higher than that of the Cu 3$d_{x^2-y^2}$ orbital in the cuprates. Thus, we should properly choose elements in the 1-3 groups (such as Sr and La) that strongly favor closed-shell electronic configuration. Following this strategy, the dynamical stability of 57 materials was examined in Ref.~\onlinecite{Hirayama_2020}. While 16 compounds do not have imaginary phonon modes, we hereafter focus on RbCa$_2$NiO$_3$ and $A_2$NiO$_2$Br$_2$ as representative compounds.

\begin{figure}
\includegraphics[width=8cm,bb=0 0 475 513]{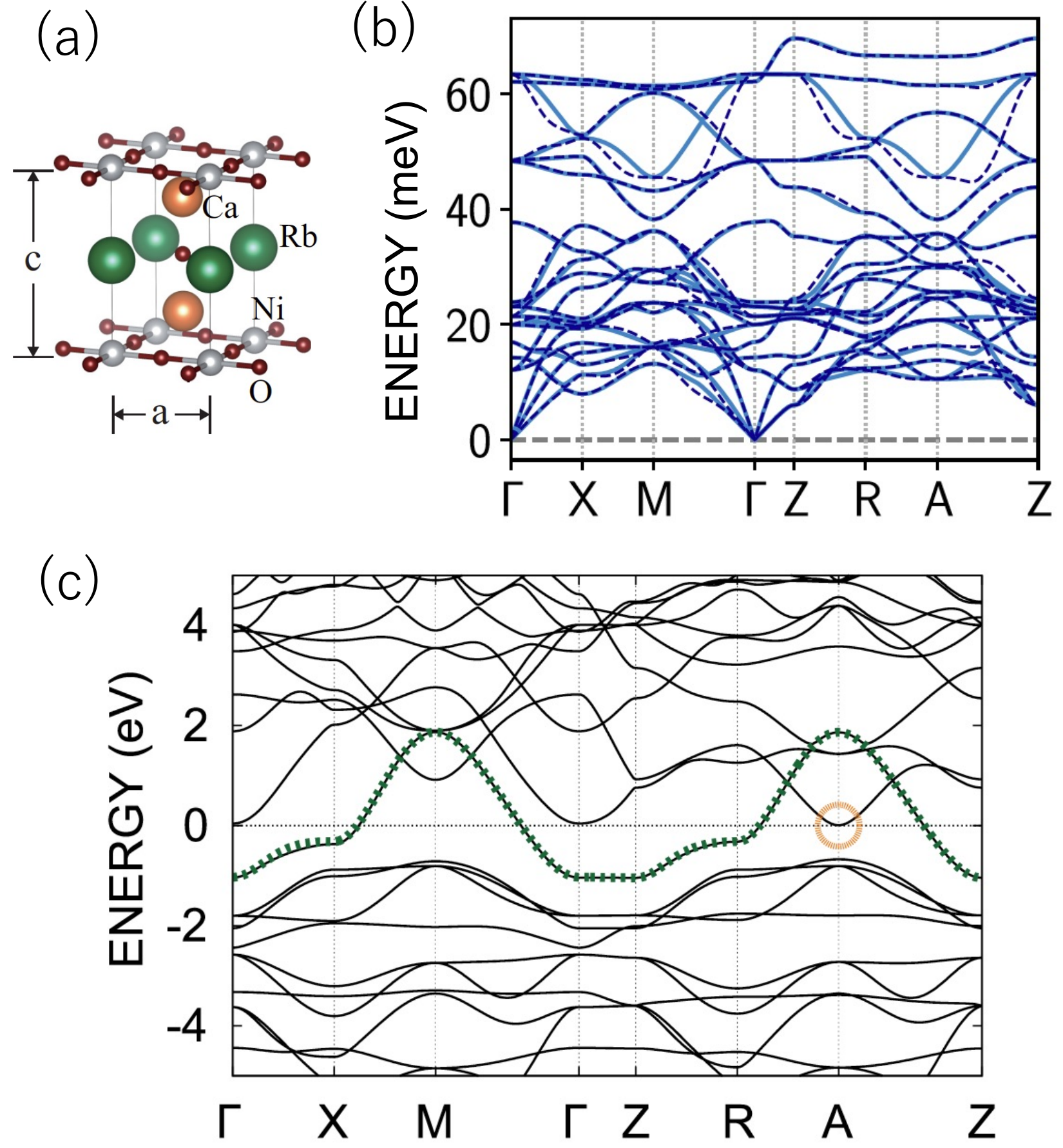}
\caption{(a) Crystal structure, (b) phonon dispersion, and (c) electronic structure of RbCa$_2$NiO$_3$. For the phonon calculation, the results for 2$\times$2$\times$2 and 4$\times$4$\times$2 supercells are shown with solid and dashed lines, respectively. The green dotted curves in (c) is the band dispersion of the single-orbital model. The open circle indicates the band minimum of the block-layer band. Taken from Ref.~\onlinecite{Hirayama_2020} and Ref.~\onlinecite{Nomura_2020}.}
\label{Fig2}
\end{figure}

\begin{figure}
\includegraphics[width=8cm,bb=0 0 480 533]{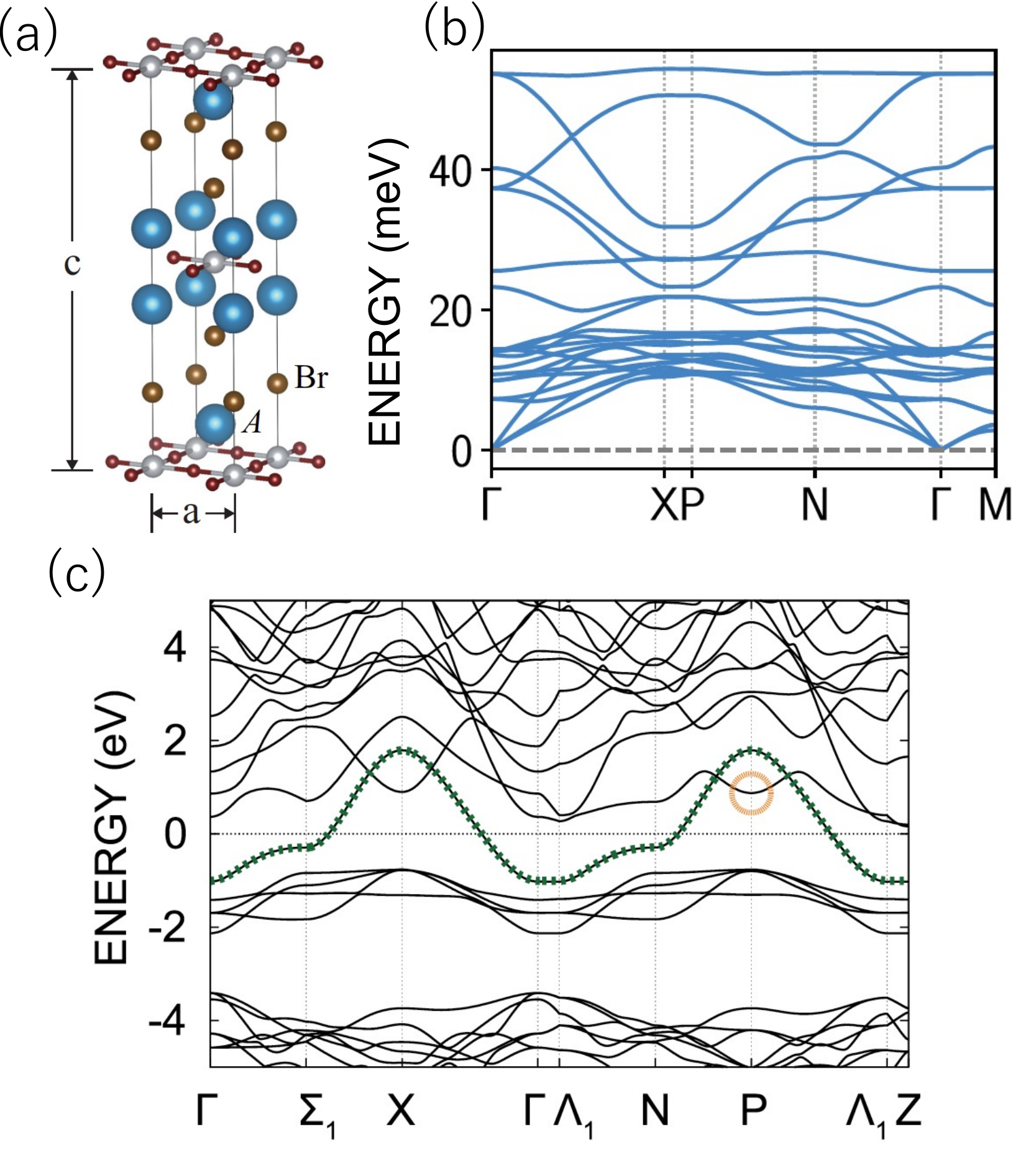}
\caption{(a) Crystal structure, (b) phonon dispersion, and (c) electronic structure of $A_2$NiO$_2$Br$_2$ ($A$ denotes a cation with a valence of 2.5+). For the phonon calculation, a 2$\times$2$\times$2 supercell was used. The green dotted curves in (c) is the band dispersion of the single-orbital model. The open circle indicates the band minimum of the block-layer band. Taken from Ref.~\onlinecite{Hirayama_2020} and Ref.~\onlinecite{Nomura_2020}.}
\label{Fig3}
\end{figure}

In Fig.~\ref{Fig2} and Fig.~\ref{Fig3}, we show the crystal structure, phonon dispersion, and electronic structure of RbCa$_2$NiO$_3$ and $A_2$NiO$_2$Br$_2$, respectively. For the calculation of the phonon dispersion, the frozen-phonon method with a 2$\times$2$\times$2 or 4$\times$4$\times$2 supercell was employed. We see that calculation for the 2$\times$2$\times$2 supercell is enough to examine the presence/absence of imaginary modes. Since the calculation of the convex hull of ternary compounds is extremely expensive, we will not discuss whether these compounds are thermodynamically stable. However, from the results in Fig.~\ref{Fig2}(b) and Fig.~\ref{Fig3}(b), we can safely conclude that they are at least dynamically stable. Here, it is interesting to note that there is an experimental report of the synthesis of the $d^8$ nickelate Sr$_2$NiO$_2$Cl$_2$~\cite{Tsujimoto_2014}, which has the same crystal structure as $A_2$NiO$_2$Br$_2$. The phonon band width is about 60-70 meV, which is similar to that of NdNiO$_2$ (see Fig.~\ref{Fig1}).

In Fig.~\ref{Fig2}(c) and Fig.~\ref{Fig3}(c), we show the electronic structure of RbCa$_2$NiO$_3$ and $A_2$NiO$_2$Br$_2$, respectively. Starting from these results, one can derive a single-orbital tight-binding model for the Ni 3$d_{x^2-y^2}$ orbital~\cite{Hirayama_2020}. We highlight the band dispersion of the effective model with green dotted curves. As we discuss in Section~\ref{sec:3}, the dispersion can be represented by a simple tight-binding model on the 2D square lattice. We also show the band minimum of the block-layer band with open circles. We see that the band minimum is higher than the Fermi level, so that the charge transfer from the NiO$_2$ to the block layer does not occur. Therefore, the effective Coulomb interaction between the Ni 3$d_{x^2-y^2}$ electrons tends to be stronger than that in $R$NiO$_2$ since the screening from the block layer becomes weaker.
%the block layer does not contribute to metallic screening.

\section{Effective low energy models\label{sec:3}}
In this section, we look into the detail of the effective single-orbital model whose Hamiltonian is given by
\[
{\mathcal H}=\sum_{ij\sigma}t_{ij}(c^\dagger_{i\sigma}c_{j\sigma}+h.c.)+U\sum_i n_{i\uparrow}n_{i\downarrow},
\]
where $c^\dagger_{i\sigma}$ and $c_{i\sigma}$ are the creation and annihilation operator for the Ni 3$d_{x^2-y^2}$ orbital at site $i$ with spin $\sigma$. $n_{i\sigma}=c^\dagger_{i\sigma}c_{i\sigma}$ is the density operator. The transfer integrals $\{ t_{ij}\}$ between site $i$ and $j$ can be evaluated by calculating the matrix elements of the Kohn-Sham Hamiltonian (${\mathcal H}_{\rm KS}$) in terms of the Wannier functions $\{|\phi_{i\sigma}\rangle\}$ for the Ni 3$d_{x^2-y^2}$ orbital,
\[
t_{ij}=\langle \phi_{i\sigma}| {\mathcal H}_{\rm KS}|\phi_{j\sigma}\rangle.
\]

On the other hand, the cRPA~\cite{Aryasetiawan_2004} is used to evaluate the effective Coulomb interaction in the model. It should be noted that the screening effect of the Ni 3$d_{x^2-y^2}$ electrons should be considered when we solve the effective model, and we should not take account of that effect when we derive the model. For this purpose, in the cRPA calculation, we first divide the polarization function $P$ into the two parts, $P_l$ and $P_h$. Here, $P_l$ includes the contribution of the transition processes between the the Ni 3$d_{x^2-y^2}$ states, and $P_h$ is the other contributions to $P$. We then calculate
\[
W_h=[1- \varv P_h]^{-1} \varv,
\]
where $\varv$ is the bare Coulomb interaction. Since
\[
W=[1-\varv P]^{-1}\varv=[1-W_hP_l]^{-1}W_h,
\]
we see that $W_h$ plays the role of the bare Coulomb interaction in the subspace of the Ni 3$d_{x^2-y^2}$ states. Taking the value of $W_h$ in the limit of the zero frequency, the Hubbard $U$ can be calculated as follows:
\[
%U=\langle \phi_{i\sigma} \phi_{i\sigma} |W_h|\phi_{i\sigma}\phi_{i\sigma}\rangle.
\cyan{U=\langle \phi_{i\sigma} \phi_{i\sigma} |W_h(\omega =0)|\phi_{i\sigma}\phi_{i\sigma}\rangle.}
\]

In Table~\ref{model}, we list the values of the transfer integrals between the nearest neighbor sites ($t$), next nearest neighbor sites ($t'$), third nearest neighbor sites ($t''$), and Hubbard $U$~\cite{Nomura_2019,Hirayama_2020, Nomura_2020}. We see that the ratio between $U$ and $t$ in RbCa$_2$NiO$_3$ and $A_2$NiO$_2$Br$_2$ is substantially larger that in NdNiO$_2$.

\begin{table} 
\caption{
Hopping and interaction parameters in the single-orbital Hubbard model. 
$t$, $t'$, $t''$ are the nearest, next-nearest, and third-nearest hopping integrals, respectively. 
$U$ is the onsite Hubbard interaction. 
The energy unit is eV.
Taken from Ref.~\onlinecite{Nomura_2019}, Ref.~\onlinecite{Hirayama_2020} and Ref.~\onlinecite{Nomura_2020}.
} 
\vspace{0.2cm}
%{\scriptsize %%%%%%%%%%%%%%%%%%%%
%\begin{tabular}{@{\ \ \ \ }c@{\ \ \ \ \ \ \ }c@{\ \ \ \ \ }c@{\ \ \ \ \ }c@{\ \ \ \ \ }c@{\ \ \ \ \ }c@{\ \ \ \ \ }c@{\ \ \ \ }}
\begin{tabular}{@{\ \ \ \ }c@{\ \ \ \ \ \ \ }c@{\ \ \ \ \ }c@{\ \ \ \ \ }c@{\ \ \ \ \ }c@{\ \ \ \ \ }c}
\hline \hline \\ [-8pt]   
 & $t$  & $t'$ & $t''$ & $U$ & $|U/t|$   \\ [+1pt]
\hline \\ [-8pt] 
NdNiO$_2$       & $-0.370$ & 0.092 & $-0.045$ & 2.608  & 7.052  \\ [+1pt]
RbCa$_2$NiO$_3$ & $-0.352$ & 0.100 & $-0.046$ & 3.347 & 9.522  \\ [+1pt]
$A_2$NiO$_2$Br$_2$ & $-0.337$  &  0.089 &  $-0.039$ & 3.586 & 10.637   \\ [+1pt]
\hline \hline 
\end{tabular}
%} %%%%%%%%%%%%%%%%%%
\label{model} 
\end{table}

\section{D$\Gamma$A calculation and phase diagram\label{sec:4}}
%Kitatani-sama: (1) what is D$\Gamma$A. (2) Result (phase diagram). (3) Brief Discussion.
%Comparison with recent Kitatani-san's calculation~\cite{Kitatani_2020,Worm_2021}.

Next, let's analyze the correlation effect of these effective low energy models for studying the superconductivity property. Based on Table I, we studied the two dimensional Hubbard model with hoppings: $t^{\prime}/t=-0.28,\; t^{\prime\prime}/t=0.13,\; U=9.5t$ for RbCa$_2$NiO$_3$ and $t^{\prime}/t=-0.26,\; t^{\prime\prime}/t=0.12,\; U=10.5t$ for A$_2$NiO$_2$Br$_2$. Here, we employ the D$\Gamma$A~\cite{Toschi_2007,Katanin_2009,Rohringer_2018}, which is one of the diagrammatic extensions of the dynamical mean field theory (DMFT)~\cite{Metzner_1989,Georges_1992,Georges_1996}. The D$\Gamma$A can capture the strongly correlated effect and the long-range fluctuation effect beyond DMFT, both of which are essential for describing the layered unconventional superconductivity \mk{with strong correlations}. The method has succeeded in describing the phase diagram of the unconventional superconductivity~\cite{Kitatani_2019,Kitatani_2020}, e.g., cuprates and nickelates. In calculation details, we follow the previous paper about the infinite layer nickelates \cite{Kitatani_2020}. Please see that paper or review articles~\cite{Rohringer_2018,Held_2022,Kitatani_2022} for more information about the method.

\begin{figure}
\centering
\includegraphics[width=\linewidth,angle=0]{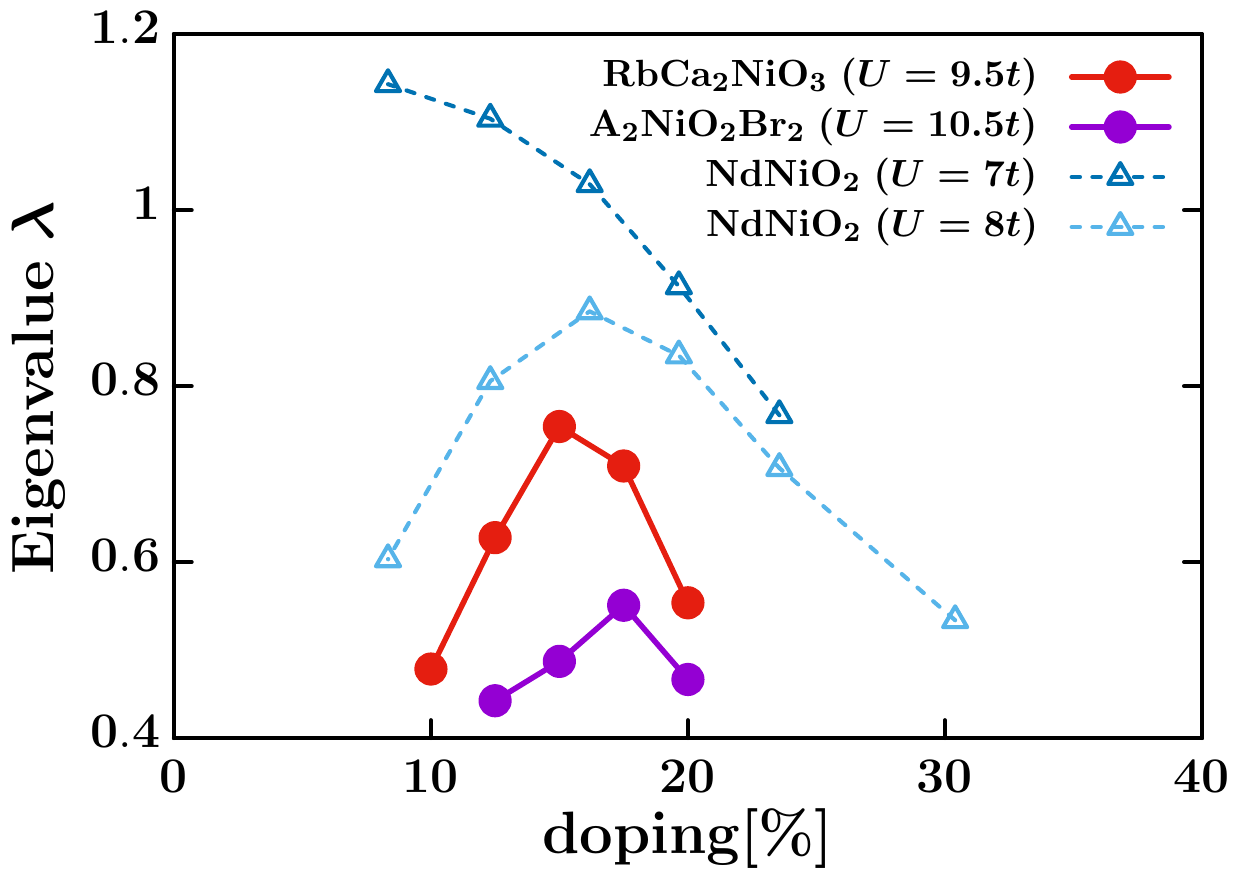}
\caption{Leading superconductivity eigenvalues for NdNiO$_2$, RbCa$_2$NiO$_3$, and A$_2$NiO$_2$Br$_2$ at fixed temperature $T=0.01t$. NdNiO$_2$ results are taken from Ref.\onlinecite{Kitatani_2020}.}
\label{fig:phase}
\end{figure}

\begin{figure}
\centering
\includegraphics[width=\linewidth,angle=0]{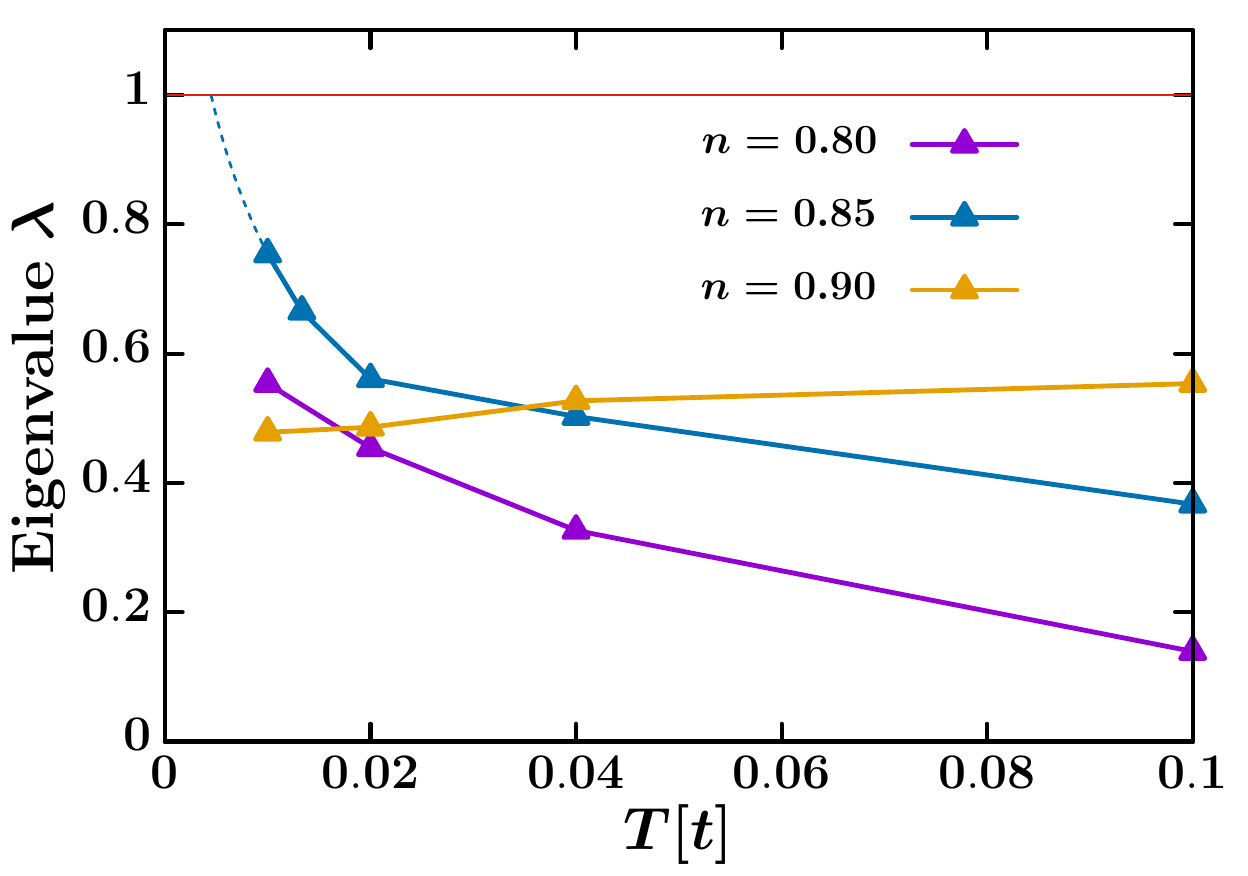}
\caption{Temperature dependence of the superconductivity eigenvalue for RbCa$_2$NiO$_3$.}
\label{fig:lambda-T}
\end{figure}

In Fig.~\ref{fig:phase}, we show the D$\Gamma$A result of leading eigenvalues of the linearized gap function in the $d$-wave (singlet, even-frequency) channel \mk{against the hole doping level}. Here, we also show NdNiO$_2$ results taken from Ref.~\onlinecite{Kitatani_2020}, where the charge transfer effect is also taken into account. The eigenvalue is often used as the measure of $T_{\rm c}$ since it usually monotonically increases as decreasing the temperature and the phase transition occurs when the eigenvalue reaches the unity.
We can see that both RbCa$_2$NiO$_3$ and A$_2$NiO$_2$Br$_2$ show the dome structure of the superconductivity instability peaked at around 15-20\% doping like cuprates and the infinite-layer nickelates. On the other hand, eigenvalues are lower than the infinite-layer system\mk{~\cite{Kitatani_2020} or the quintuple-layer system~\cite{Worm_2021}}. 
Some theoretical studies suggest that the infinite-layer nickelate material resides in the strong-coupling regime of superconductivity~\cite{Sakakibara_2020,Kitatani_2020}. Indeed, the $T_{\rm c}$ change in recent pressure and strain experiments~\cite{Ren_2021,NN_Wang_2021,Lee_2022} can be reasonably understood by the change of the effective interaction strength.
Both materials considered here have further large $U/t$ values, and then we obtained weaker superconductivity instability.

We also show the temperature dependence of eigenvalues for RbCa$_2$NiO$_3$ in Fig.~\ref{fig:lambda-T}. At the optimal doping regime at $n=0.85$, we obtained the transition temperature $T_{\rm c} \sim 0.0045t \sim 18$ K by extrapolating \mk{the D$\Gamma$A result} with the fit function $\lambda(T)=a-b \ln{(T)}$~\cite{Sekine_2013,Kitatani_2020}. Here we just used the $U/t$ value obtained from cRPA calculations. If we employ a somewhat larger $U/t$ value like Ref. \onlinecite{Kitatani_2020}, $T_{\rm c}$ will further decrease.

\begin{figure}
\centering
\includegraphics[width=\linewidth,angle=0]{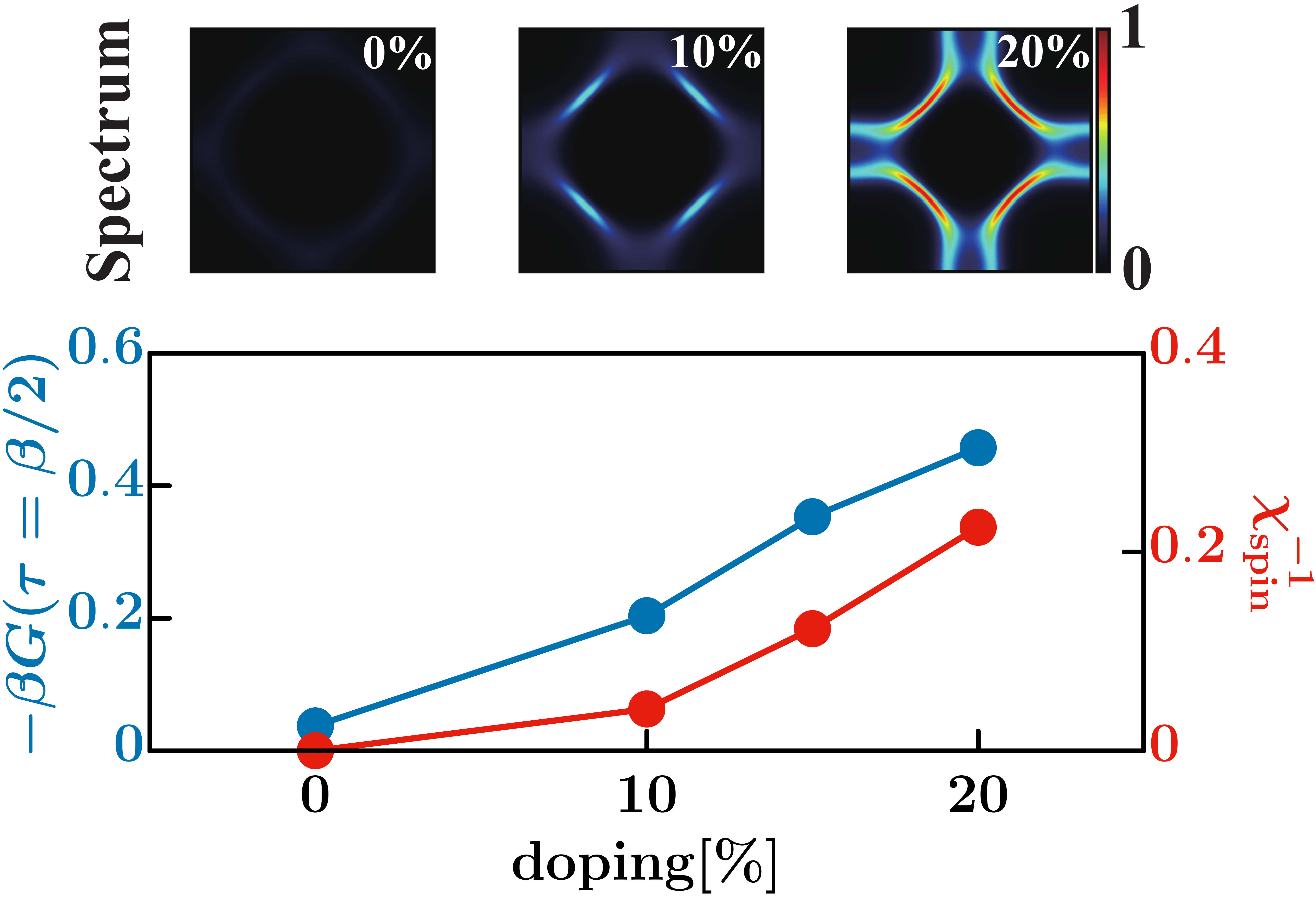}
\caption{Filling dependence of the spectrum (Green function) and the spin susceptibility $\chi_{\rm spin}({\bm Q}_{\rm max},\omega=0)$ for RbCa$_2$NiO$_3$ ($t^{\prime}/t=-0.28, t^{\prime\prime}/t=0.13$) at $T=0.04t$.}
\label{fig:fillingdep}
\end{figure}

Furthermore, we show the filling dependence of the spectrum and spin fluctuation in Fig. \ref{fig:fillingdep}. \mk{As a proxy of the spectrum, we here show the momentum dependence of the imaginary part of the Green function at the lowest Matsubara frequency $-\frac{1}{\pi}G({\bm k},\omega_n=\pi/\beta)$ and the momentum averaged Green function at $\tau=\beta/2$.} Without doping, we can see the similarity with cuprates: the spectral weight at the Fermi level is strongly suppressed by the self-energy, and spin fluctuation becomes strongly enhanced so that antiferromagnetism will stabilize once we consider the weak three dimensionality.

\section{Conclusion\label{sec:5}}
In conclusion, we performed a calculation based on the dynamical vertex approximation and obtain the phase diagram of RbCa$_2$NiO$_3$ and $A_2$NiO$_2$Br$_2$. In these materials, the charge transfer from the NiO$_2$ layer to the block layer is absent, and the effective Coulomb interaction between Ni 3$d_{x^2-y^2}$ electrons is stronger than that in $R$NiO$_2$ or Nd$_6$Ni$_5$O$_{12}$. \mk{We obtained the dome shaped superconductivity instability like infinite layer nickelates. While the estimated $T_{\rm c}$ is lower than $R$NiO$_2$~\cite{Kitatani_2020}, we here demonstrate that we can control the charge transfer and correlation effect among nickelate compounds by changing block layers.}  Furthermore, RbCa$_2$NiO$_3$ and $A_2$NiO$_2$Br$_2$ may provide us with a unique opportunity to study the electron-hole asymmetry of the Mott insulating states in the nickelate superconductors.

\begin{acknowledgments}
We are grateful for fruitful discussions with Karsten Held, \mk{Liang Si, Paul Worm,} Terumasa Tadano, Takuya Nomoto, and Kazuma Nakamura. 
We acknowledge the financial support by Grant-in-Aids for Scientific Research (JSPS KAKENHI) [Grant No. \mk{JP20K22342 (MK), JP21K13887 (MK),} JP20K14423 (YN), JP21H01041 (YN), and JP19H05825 (RA)] and MEXT as ``Program for Promoting Researches on the Supercomputer Fugaku'' (Basic Science for Emergence and Functionality in Quantum Matter ---Innovative Strongly-Correlated Electron Science by Integration of ``Fugaku'' and Frontier Experiments---) (JPMXP1020200104). \mk{A part of the calculation has been done on MASAMUNE-IMR of the Center for Computational Materials Science, Tohoku University.}
\end{acknowledgments}

\section*{Data Availability Statement}
The data that support the findings of this study are available within the article.

%\nocite{*}
\bibliography{nickelate}% Produces the bibliography via BibTeX.

\end{document}